**Active Microrheology of Networks Composed of Semiflexible Polymers**

**I. Computer Simulation of Magnetic Tweezers**

N. Ter-Oganessian*, D. A. Pink♦, B. Quinn♦, E. and A. Boulbitch*

*Dept. for Biophysics E22, Technical University Munich, James-Franck-Str. 1,

D-85747 Garching, Germany

♦ Physics Department, St Francis Xavier University, Antigonish, Nova Scotia, Canada B2G2W5

**Abstract**

We have simulated the motion of a bead subjected to a constant force while embedded in a network of semiflexible polymers which can represent actin filaments. We find that the bead displacement obeys the power law $x \sim t^{\alpha}$. After the initial stage characterized by the exponent $\alpha_1 \approx 0.75$ we find a new regime with $\alpha_2 \approx 0.5$. The response in this regime is linear in force and scales with the polymer concentration as $c^{-1.4}$. We find that the polymers pile up ahead of the moving bead, while behind it the polymer density is reduced. We show that the force resisting the bead motion is due to steric repulsion exerted by the polymers on the front hemisphere of the bead.

# 1. Introduction

## 1.1. Microrheology

In animal cells the space between the cell nucleus and the membrane is occupied by the cytoskeleton. It represents a network composed of several types of filamentous proteins and enables cells to bear and respond to mechanical loads[1]. The filamentous protein, F-actin, is the major component of the cytoskeleton which explains its extraordinary role in cell mechanics and the interest in its mechanical properties[2, 3]. The mechanical behavior of actin filaments in solution has been studied experimentally by several techniques such as dynamic light scattering[4], macrorheology[5], and magnetic tweezers[6-12] to quantify its viscoelastic properties.

In such measurements one distinguishes between passive and active approaches. Passive measurements are defined by analyzing the thermal fluctuations of beads[3, 13, 14], while, during active microrheological experiments, a force is applied to the bead by a laser beam (optical tweezers) [15, 16] or by magnetic field (magnetic tweezers, requiring paramagnetic microbeads)[3] and the displacement of the bead is measured. It is important to understand local (i.e. within a few micrometers) viscoelastic properties of an actin network, since it is on this scale that mechanical loads in the range 10 to $10^6$pN often act on cells *in vivo*[17-20]. This requires using active microrheology and is achieved by a technique based on microbeads which has been widely used as a tool to probe local microrheological properties of biological materials [6, 7, 21-31].

In this paper we present a model of a microbead embedded in an aqueous solution containing semiflexible polymers viewed as a network of actin filaments. We shall refer to models of semi-flexible linear macromolecules as "polymers". In accord with current usage, we use "filaments" to describe macromolecules of actin. We consider a constant force applied to the microbead and we search for the dependence of microbead dynamics upon the parameters of the system. Because of the complexity of the model we resort to computer simulation. In creating the model, we recognize that the key elements are: a dense solution of polymers in water in which a bead of dimensions much larger than the mesh size is embedded. We utilize a minimal model of this system, representing the polymers as beads connected by springs and making use of dissipative particle dynamics. Our intent is to simulate experiments reported elsewhere[32, 33] and to understand a possible mechanism for the newly observed regime. An analytic theory describing microrheological measurements will be reported in a forthcoming paper[34]. In what follows we briefly review experimental results[32, 33].

The magnetic tweezers approach was introduced in 1922[21] and widely used to study micromechanical properties of materials as different as the human vitreous body[26], various cells[23, 24, 27-30, 35-37] and actin networks[6-12, 31, 38, 39]. The magnetic tweezers technique consists of embedding a paramagnetic bead in the medium and applying to it an inhomogeneous magnetic field $\mathbf{B}$. The effect of the field is to (a) create a magnetic moment in the bead in the direction of $\mathbf{B}$ and (b) set up a force $\mathbf{f} \sim (\mathbf{B} \cdot \nabla)\mathbf{B}$ acting on the bead without applying any torque[6-12]. Both the force $f$ applied to the bead as well as the bead displacement $x = x(t)$ are measured[6, 7, 11].

### 1.2. Microrheology of Actin Networks: Recent Findings

Recently the time resolution of magnetic tweezer measurements was improved 10-fold and force pulses of up to 60s duration were applied[32, 33]. The displacement of a bead in a viscoelastic medium can be described in terms of a compliance, $J(t)$, defined by[14] $x(t) = J(t)f / 6\pi R_b$, where $R_b$ is the bead radius. Measurements[32, 33] revealed that the compliance of a tightly entangled actin network obeyed the relation $J(t) \approx A_i t^{\alpha_i} + B_i$ over five decades ( $0.6\text{ms} \leq t \leq 60\text{s}$ ) and exhibited three regimes, labeled by $i = 1, 2, 3$, characterized by the exponents $\alpha_i$, the amplitudes, $A_i$ and the offsets $B_i$ with $B_1 = B_2 = 0$. For the intermediate-time regime, $i=2$ (which we address in this work), it is convenient to write the time-dependence of bead displacement as

$$x = Kt^{1/2} \tag{1}$$

$K$ is related to the compliance coefficient $A_2$ by $K = A_2 f / 6\pi R_b$.

During the initial regime $\tau_i < t < \tau_1$ the data can be fitted by $\alpha_1 \approx 0.75$, where $\tau_i \approx 0.6\,\mathrm{ms}$ is the time resolution of the set up and the crossover time $\tau_1$ varies between 0.05 and 0.3s[32, 33]. The exponent value $\alpha_1 \approx 0.75$ is in accord with earlier observations[40] and agrees with the high frequency dependence of the shear modulus[41, 42] $G \sim \omega^{3/4}$.

The intermediate-time regime of the bead motion was observed for $\tau_1 < t < \tau_2$, where the crossover time $\tau_2$ varies between 10 and 30s depending on the force applied to the bead. In this interval the power law $x \sim t^{\alpha_2} c^{-\gamma_2}$, with $\alpha_2 \approx 0.5$ and $\gamma_2 \approx 1.1$ was observed to hold over two decades.

At $10\mathrm{s} \leq t < 60\mathrm{s}$ (the long-time regime) the exponent $\alpha_3$ was observed to increase to 0.9, probably indicating a crossover to the viscous regime in which $\alpha_3 = 1$[32, 33].

Several theoretical models have been proposed to describe the viscoelastic behavior of semi-dilute actin solutions[3, 13, 14]. Progress has been achieved in understanding the mechanism of the high-frequency dependence of the shear modulus[42] $G(\omega) \sim \omega^{3/4}$ as being due to the bending[40] and stretching[41] of the filaments. The reptation tube model[43-46] was used to account for its low-frequency viscoelastic behavior. The model accounting for the diffusion of the excess lengths of the filaments along the reptation tubes predicts that $G(\omega) \sim \omega^{1/2}$ in the flexible limit $L_p / L << 1$. This relation however, does not hold for semiflexible polymers with $L_p / L \sim 1$, where $L_p$ is the persistent and $L$ is the contour length of the polymer[44]. A phenomenological model describing small bead displacements based on two-fluid hydrodynamics has been developed[47, 48]. However, these models do not describe the square-root regime observed in the case of the semiflexible actin filaments[32, 33].

In this paper we model the enforced motion of a bead through a network consisting of a tightly-entangled solution of semiflexible polymers and we study its dynamics by simulating it using the Dissipative Particles Dynamics (DPD) method[49, 50]. The paper is organized as follows. In Section 2 we briefly describe the DPD method and our model. Details of the DPD equations are given in Appendix A, while the choice of the model parameters is found in Appendix B. Section 3 contains the results of our simulations while in Section 4 we discuss and analyze them. The conclusions are in Section 5.

On the basis of our simulations an analytical model will be proposed in the forthcoming paper[34].

## 2. Simulations

### 2.1. Dissipative Particle Dynamics

The DPD technique[49] is a model for the simulation of the hydrodynamic behavior of a fluid and bridges the gap between microscopic models and macroscopic approaches involving the solution of the fluid flow equations. In it the system is represented, in most cases, by point objects ('particles') possessing masses and interaction radii outside of which the interaction in question is zero. In cases where each particle possesses a unique interaction radius these interaction radii can be thought of as the effective radii of the particles. An obvious exception is the case of a short-range interaction together with electrostatics which is not treated in this paper. This method can be understood as a coarse-graining of the fluid particles on the smallest physically significant length scale so that all smaller scale motions are ignored or interpreted stochastically[51]. The model can be considered as a generalization of the Langevin approach and satisfies the Navier-Stokes equations[52-54] on length scales larger than the particle interaction range. The method employs molecular dynamics in the presence of conservative, random and dissipative forces. Shortly after its creation the DPD method was recognized to be well-suited for the simulation of polymer systems[55 56, 57 58 59].

In our model we utilized both water and monomer particles. The latter are connected to represent polymers (see below). In DPD all interactions are pairwise and comprise a conservative force $\mathbf{F}^{(C)}$, a dissipative force $\mathbf{F}^{(D)}$ and a random force $\mathbf{F}^{(R)}$. All interaction ranges (except those between the monomers belonging to the same polymer to be introduced below) are defined by a distance $R_\alpha$ characteristic of the particle of type $\alpha$. Here and in the following equations, the Greek subscripts indicate the type of the particles (i.e. the bead, the water sphere or the monomer) and should not be mixed up with those in the exponents denoting power laws. The conservative force is a repulsion acting along the line joining the centers of two particles. These forces are described in Appendix A while the choice of parameters of the system including the constants of the DPD forces is summarized in Appendix B.

In hydrodynamics a flow around a spherical bead is determined by fixing an appropriate boundary condition on its surface. Usually no-slip boundary conditions are used[60]. Accounting for such conditions is complicated both in simulations and in analytical calculations, and approximations are often used in practice[24, 47]. In order to simplify the algorithms in the simulations reported here we modeled the bead as a DPD particle with the radius $R_b = 10\ \mathrm{LU}$, mass $m_b$ and interaction parameters with water, $a_{bw}$, and with monomers, $a_{bm}$. The mass $m_b$ is much larger than $m_w$ and $m_m$, and the interaction constants are larger than those for water-water and monomer-monomer interactions (Table 1) by an order of magnitude. To make sure that such a model results in a correct description of the motion of a fluid around the bead we simulated the bead moving through pure water (i.e. water without polymers) and compared the results with the known solutions of Stokes equation[60]. This is described in detail in Appendix C.

In this Section we give only details concerning forces acting within a polymer.

### 2.2. DPD Model of Semiflexible Polymers

In our case, we used a bead-and-spring model of polymers composed of a sequence of connected monomer particles. Two such monomers of type $\alpha$ adjacent to one another along a polymer chain and a distance $r_{\alpha\alpha ij}$ apart interact via a harmonic potential so that the force, $F_{\alpha\alpha ij}^{(S)}$,

**Table 1**

Parameters of the DPD method

(All parameters are given in corresponding units described in the text)

| Parameter | $L_x$ | $L_y$ | $L_z$ | $R_b$ | $R_m$ | $R_w$ | $\rho$ |
|-----------|-------|-------|-------|-------|-------|-------|--------|
| Value | 40 | 80 | 40 | 10 | 0.3 | 0.8 | 3 |
| Parameter | $a_{ww}$ | $\gamma_{ww}$ | $a_{mm}$ | $\gamma_{mm}$ | $a_{mw}$ | $\gamma_{mw}$ | $\delta_{NS}$ |
| Value | 45 | 1 | 35 | 5 | 0 | 1 | 1 |
| Parameter | $a_{bw}$ | $\gamma_{bw}$ | $a_{bm}$ | $\gamma_{bm}$ | $k_B T$ | $\Delta t$ | $L$ |
| Value | 550 | 0.1 | 550 | 0.1 | 1 | $5 \times 10^{-5}$ | 34.5 |
| Parameter | $m_b$ | $m_w$ | $m_m$ | $d$ | $k$ | $\mu$ | $N_{m/p}$ |
| Value | 1 | 0.0001 | 0.0001 | 0.5 | 400 | 385 | 70 |

acting on particle $\alpha i$ due to the adjacent particle $\alpha j$ is:

$$\mathbf{F}_{\alpha\alpha ij}^{(S)} = -k(r_{\alpha\alpha ij} - d)\hat{\mathbf{r}}_{\alpha\alpha ij}, \qquad (2)$$

where $k$ is the spring constant, $d$ is the equilibrium length of the monomer-monomer bond and $\hat{r}_{\alpha\alpha ij}$ is a unit vector pointing from $\alpha j$ to $\alpha i$.

To model a semiflexible, rather than a flexible polymer we introduced "persistence" forces, which straighten the polymers (Fig. 1 (a)) and are defined according to relations

$$F_n = \mu\varphi / r_n \qquad (3)$$

where $n = 1$ or $2$, and $r_1$ and $r_2$ are the lengths of the bonds 1-3 and 2-3, respectively (Fig. 1 (a)), $\varphi$ is the complementary angle between these bonds and $\mu$ is the force constant. The force $\mathbf{F}_n$ acts perpendicular to the bond of length $r_n$ (n = 1,2) and is chosen to lie in the plane formed by the triangle 1-2-3. We define the force $\mathbf{F}_3 = -\mathbf{F}_1 - \mathbf{F}_2$ to be acting on the monomer 3 (Fig. 1 (a)) so

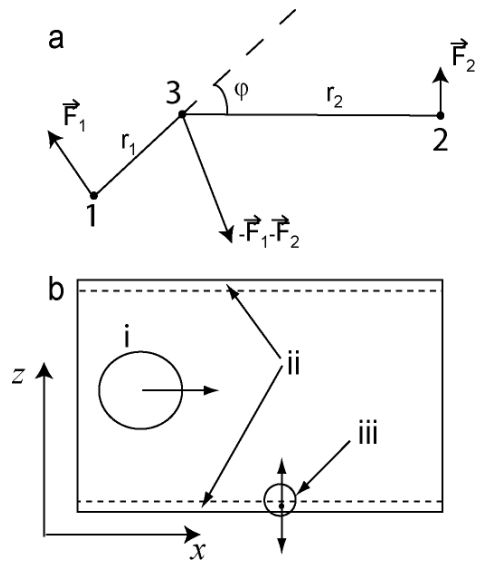

Fig. 1  (a) Persistence forces. The forces $\mathbf{F}_1$ and $\mathbf{F}_2$ act on the monomers 1 and 2 and are normal to the bonds 1-3 and 2-3, respectively. The force $\mathbf{F}_3$ acting on the monomer 3 is opposite to the vector sum $\mathbf{F}_1 + \mathbf{F}_2$. All forces lie in the plane defined by the monomers 1, 2 and 3.  (b) Thin non-slip layers near the walls parallel to the direction of the application of the external force on the bead. The water spheres in the layers are only able to move perpendicular to them. (i) the bead, the (ii) the boundary layer in which the non-slip conditions are imposed and (iii) the water sphere at the boundary.

that both the total force and the torque acting on the trimer is zero. Such a choice of forces favors the straight configuration with $\varphi = 0$ thus enabling us to choose a persistence length of the polymers.

We confirmed via simulations that our choice of polymer parameters resulted in a persistence length that reflected a ratio of bead size to persistence length appropriate to the case of actin filaments. This is described in more detail in Appendix B.

## 2.2. Simulation of the bead motion through the polymer network

We performed computer simulations of a system composed of water, semiflexible polymers and a bead as described above. The choice of the parameters is described in Appendix B and summarized in Table 1. The system was simulated in a 3D box with dimensions $L_x$, $L_y$ and $L_z$ measured in arbitrary length units (referred to as LU) and satisfying periodic boundary conditions. All polymers had the same contour length $L$ which was chosen to be smaller than the smallest dimension of the simulation box in order to prevent any artificial self-interaction of the polymers across the box walls.

We simulated four systems containing $N_1 = 2900$, $N_2 = 3500$, $N_3 = 4000$ and $N_4 = 4500$ polymers. The mesh sizes of these networks, $\xi_i$, can be estimated according to the relation $V = gN_iL\xi_i^2$, where $N_i$ is the number of polymers, $i$=1, 2, 3 and 4 indicates the system with the corresponding number of polymers, $V = L_xL_yL_z$ is the box volume and $g$ is the geometric factor. In the case of polymers lying along a primitive cubic lattice one finds $g = 1/3$ which we will assume in the following. The concentration of polymers is defined as $c_i = N_i/V$ yielding $c_i = 3/L\xi_i^2$. The values of the corresponding mesh sizes and concentrations are summarized in the Table 2. For all these systems the inequality $\xi_i/L_p << 1$ holds, ensuring that we simulated a tightly-entangled regime.

In order that water particles not accumulate linear momentum gained by their interaction with the enforced motion of the bead, we imposed non-slip boundary conditions on those water spheres at the walls parallel to the direction, $Ox$, of the applied force. This was achieved by imposing conditions on the velocity of any water sphere whose center was located within a layer of thickness $\delta_{NS}$ (defined in Appendix B) at any one of the four following planes: the $x$-$y$ plane at $z = 0$ and at $z = L_z$ and the $x$-$z$ plane at $y = 0$ and at $y = L_y$. The conditions were that water spheres would have their velocity components parallel to the surfaces set equal to zero after every step, as long as their centers were located inside one of these layers (Fig. 1 b). However, since the size of the box is comparable to the length of the polymers, at any time one would find a large number of polymers crossing the box surfaces. If the no-slip boundary conditions were applied to polymers as well, they would become immobile. For this reason we did not apply the no-slip boundary conditions to the polymers.

The simulations were initialized in a state with zero total momentum and with a Maxwell distribution of the velocities corresponding to the temperature $T$. In all simulations we set $k_\mathrm{B}T = 1$.

The sphere representing the magnetic bead embedded in the polymer network was subjected to a constant force in the $Ox$ direction. The systems were simulated for approximately 2 to 6 time units (TU). During the simulations the bead moved through a distance comparable to its radius. Longer simulations of these systems have not yet been performed. We used the time step $\Delta t = 5 \times 10^{-5}\,\mathrm{TU}$.

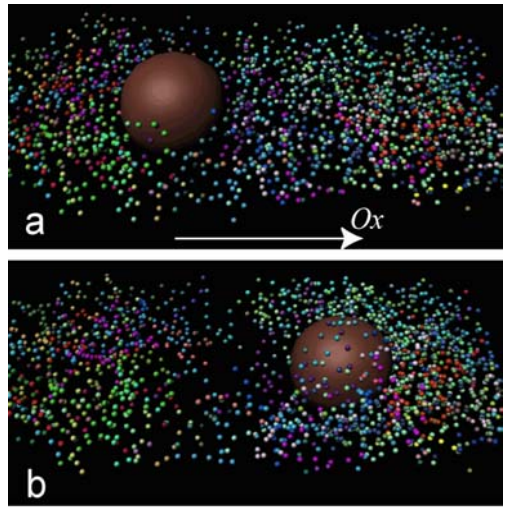

Fig. 2 Magnetic bead (brown) moving in the polymer network. The polymers are shown as small colored spheres, with beads of the same color belonging to the same polymer. For the sake of visibility we (i) show only each tenth monomer sphere, and (ii) polymers are not shown if they are located between the observer and the sphere or on the other side of the sphere from the observer. (a) The bead in a rest state (before the force is applied) surrounded by the polymers. (b) The bead during its motion along $Ox$ axis. The concentration of the polymers is increased in front of the bead and decreased in its back.

A typical view of a motionless bead with zero applied force embedded into network is shown in Fig. 2 (a). Figure 2 (b) shows its enforced motion through the network. This motion is described in the following Sections.

# 3. Results

## 3.1. Displacement of the Bead Under a Force Pulse

After the system had relaxed, a constant force was applied to the bead in the $Ox$ direction. We analyzed the bead motion in response to the forces $f_1 = 200$, $f_2 = 400$, $f_3 = 800$, $f_4 = 1000$ and $f_5 = 1300$ force units (FU) in four systems. The latter differed from one another in the numbers of polymers, as summarized in Table 2. Displacements of the bead subjected to these forces are shown in Fig. 3 (a). In Fig. 3 (b) displacements of a bead under the action of the force $f_4 = 1000\,\text{FU}$ are shown on a double-logarithmic scale for the four concentrations of Table 2. The data clearly reveals two regimes of motion:

At $t \leq 0.2\,\text{TU}$ (which corresponds to 4000 simulation steps) the fit yielded the exponent value $\alpha_1 \approx 0.75$.

**Table 2**

Numbers of polymers and corresponding values of mesh sizes and concentrations

|  | 1 | 2 | 3 | 4 |
|---|---|---|---|---|
| $N$ | 2900 | 3500 | 4000 | 4500 |
| $\xi$ (LU) | 1.96 | 1.78 | 1.67 | 1.57 |
| $c\,(\text{LU}^{-3})$ | 0.023 | 0.027 | 0.031 | 0.035 |

At $t > 0.2\,\text{TU}$ the data was fitted by the power law (1) with the exponent $\alpha_2 \approx 0.5$. This regime will be referred to as the "square-root regime". Due to time limitations on our simulations we were unable to reach the cross-over from the square-root to any other regime. Thus, for example, we did not reach the viscous regime $\alpha_3 = 1$ observed in experiments[33] and reported in[32, 33], since the regimes with $x \sim t^{0.75}$ and $x \sim t^{0.5}$ covered the total duration of our simulations.

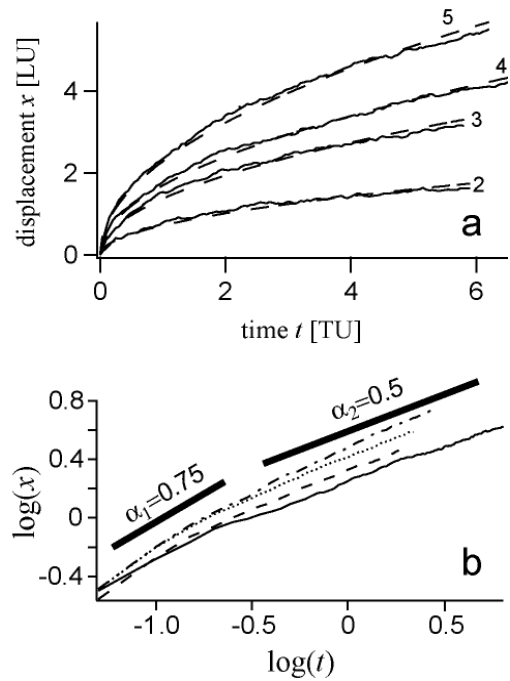

Fig. 3. (a) Typical displacements of the bead on time. Curves 2 to 5 correspond to the forces $f_2 = 400\,\text{FU}$, $f_3 = 800\,\text{FU}$, $f_4 = 1000\,\text{FU}$ and $f_5 = 1300\,\text{FU}$ applied to the bead. All curves were obtained for the system containing 4500 polymers. The dashed lines show the best fit to each curve by Eq. (1). (b) The bead displacement versus time shown on a double-logarithmic plot reveals two distinct regimes of motion: the initial regime characterized by the exponent $\alpha_1 \approx 0.75$ followed by the regime with the exponent $\alpha_2 \approx 0.5$. Thick solid lines indicate slopes with the exponents $\alpha_1 = 3/4$ and $\alpha_2 = 1/2$. Responses of the four systems are shown with various numbers of polymers: 2900 (dotted-dashed), 3500 (dotted ), 4000 (dashed), and 4500 polymers (thin solid line). All cases shown here correspond to the force of 1000FU applied to the bead.

## 3.2. Power laws

### 3.2.1. The linearity of the bead response

We analyzed the bead displacements under the application of various forces $f = 200, 400, 800, 1000$ and $1300\,\text{FU}$ and found a linear dependence on force of the coefficient $K$ describing the bead displacement in the square-root regime (1): $K \sim f$ (Fig. 4 a) yielding $x \sim f$. Our simulations did not show a sub-linear dependence of the bead displacement on force as observed in the actin network[32, 33].

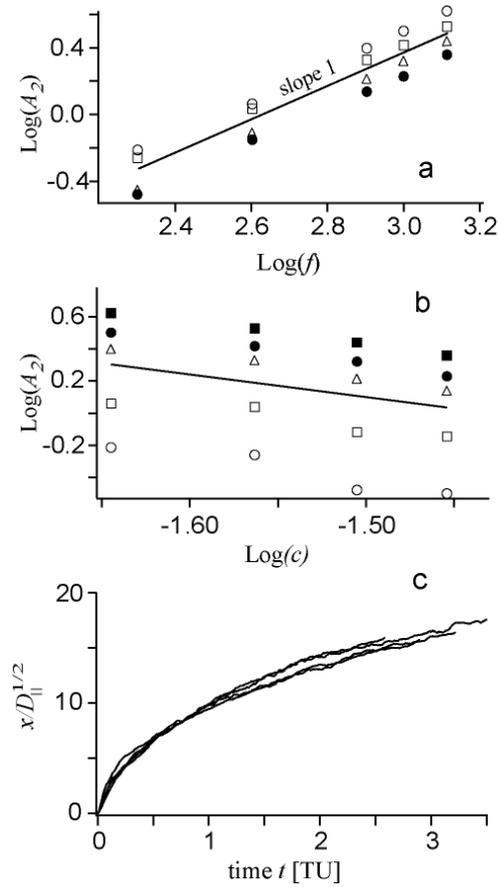

Fig. 4. (a) The coefficient $K$ characterizing the bead displacement in the square-root regime (Eq. (1)) plotted *versus* force in a double logarithmic scale exhibits a linear force dependence for the analyzed systems with 2900 (open circles), 3500 (open squares), 4000 (open triangles) and 4500 (filled spheres) polymers. The solid line indicates the slope equal to one. (b) Dependence of the coefficient $K$ on the concentration of polymers for forces $f = 200\,\text{FU}$ (open circles), $f = 400\,\text{FU}$ (open squares), $f = 800\,\text{FU}$ (triangles), $f = 1000\,\text{FU}$ (filled circles) and $f = 1300\,\text{FU}$ (filled squares). The solid line shows the slope equal to $\gamma_2 = -1.4$. (c) The bead displacements, normalized by $D_\parallel^{1/2}$ with four different dissipative constants, $\gamma_{\text{mw}}$, of monomer-water interaction, collapse onto the same line indicating the dependence $x \sim D_\parallel^{1/2}$.

### 3.2.2. Dependence of the square-root factor $A$ on the concentration of polymers

The coefficient $K$ (Eq. (1)), as well as $A_2$ depends on the number of polymers in the simulation box. For each concentration we simulated bead motion driven by five different forces, $f_1$ to $f_5$, and four concentrations of polymers, $N_1$ to $N_4$, as described in Section 2.2. We fitted the displacement, corresponding to the square-root regime, to equation (1) in order to obtain the dependence of the coefficient $K$ upon polymer concentration. The dependence upon $c$ and the fit are shown in Fig. 4 (b). Typically the displacement curves at small forces and smaller concentrations are more noisy than those in the case of higher forces. The accuracy of fitting also increases with the increase of the force and/or concentration of polymers. This analysis (Fig. 4 b) reveals the power law $K \sim c^{-\gamma_2}$ with $\gamma_2 \approx 1.4$.

### 3.2.3. Dependence of the square-root factor $A$ on the diffusion coefficient

By changing the dissipative force constant $\gamma_{mw}$ of the monomer-water interaction one can alter the longitudinal diffusion coefficient $D_\parallel$ of the simulated semiflexible polymers without altering the viscosity of the solvent or the contour length of the polymers. We performed three additional simulations of the enforced bead motion under the action of the force $f = 1000\,\mathrm{FU}$ through the polymer solution with the same mesh size $\xi = 1.57\,\mathrm{LU}$, but with different dissipative force constants $\gamma_{mw} = 0.2$, 0.4 and 0.7. The responses of the bead normalized by $D_\parallel^{1/2}$ shown in the Fig. 4 (c) fall on the same curve suggesting that $K \sim D_\parallel^{1/2}$.

## 4. Discussion

### 4.1. Comparison of predictions of our simulations with experiments

We have modeled the enforced bead motion in a network consisting of the entangled semiflexible filaments in a tightly-entangled regime (defined by the condition[43] $\xi << L_p$) in an aqueous solution. Our model, which uses dissipative particle dynamics, involved random, viscous and steric repulsion interactions between all particles as well as elastic forces between monomers belonging to the same polymers. The model involves no attractive forces and represents thus, a minimal model mimicking an entangled network consisting of semiflexible filaments.

It should be stressed that our primary interest is in $x \sim t^{1/2}$ regime of the system and we have not carried out simulations that extend to longer times in order that we get into the viscous regime. Simulations that we have carried out elsewhere (Pink, unpublished data) have shown that the effect of noise in the case of weak forces renders the identification of the various regimes difficult. Finally, we have restricted our range of polymer concentration because we wish to address sufficiently-dense cases in which the bead cannot squeeze through the mesh.

Our simulation revealed two distinct regimes of bead motion under the application of a constant external force and that could be approximately described by the power law (1). In the initial regime ($t \leq 0.2\,\mathrm{TU}$) the simulation yields the exponent $\alpha_1 \approx 0.75$ with the bead displacement in this regime being smaller than the mesh size. Evidently, this regime corresponds to the stage of motion in which the bead slightly deforms only a few polymers as already described theoretically and experimentally[40].

This corresponds to the high-frequency regime of bead fluctuations in which the shear modulus, $G$, scales as $G(\omega) \sim \omega^{3/4}$ with the frequency $\omega$. Such a behavior indeed has been observed using passive microrheology[41, 42] and in the short-time behavior of the self-displacement[61].

At later times ($t > 0.2\,\mathrm{TU}$) we observed the regime characterized by the exponent $\alpha_2 \approx 0.5$. We established that the coefficient, $K$, of the square-root regime depends upon the polymer concentration and the longitudinal diffusion coefficient, like

$$K \sim D_{\parallel}^{1/2} c^{-\gamma_2} f \qquad (4)$$

where $\gamma_2 \approx 1.4$. The square-root regime has been recently observed in the experiments on enforced motion of magnetic beads in the actin networks[33]. Our results yield dependence of $K$ on concentration, $K \sim c^{-\gamma_2}$ with $\gamma_2 \approx 1.4$ close to the value $\gamma_2 \approx 1.1 \pm 0.3$ reported by the measurements[32, 33]. In contrast, the dependence of the bead displacement on force was measured to be slightly sub-linear[33], which differs from the linear dependence (4) predicted by our simulations.

## 4.2. Distribution of polymers around the moving bead

Simulations show that during the bead motion the distribution of polymers around the bead becomes inhomogeneous. In Fig. 2 (a) the polymers in the vicinity of the motionless bead are shown while Fig. 2 (b) displays those around the moving bead. It can be seen that the polymers are piled up in front of the bead while behind the bead a region appears that is almost free of polymers.

To check this quantitatively we analyzed the distribution of polymers in the vicinity of the moving bead. During the simulations the numbers of monomers neighboring the front and the rear hemispheres of the bead were separately stored. A neighboring monomer was defined as one whose center lies within a distance of $R_m + R_b$ (see Appendix A) from the center of the bead. These monomers are also all those which are interacting with the bead. In Fig. 5 (a) the numbers of monomers in front of (indicated by (i)) and behind (indicated by (ii)) the moving bead are shown.

The distribution of the polymer concentration around the bead was also monitored during the simulations. Due to the cylindrical symmetry the concentration of polymers, $c$, at a point depends only on the distance $r$ from the center of the bead to the point and on its azimuthal angle $\theta$ measured from the $Ox$ axis. We divided the interval $\theta \in [0, \pi]$ into 12 sectors and the interval $r \in [0, 20 \text{ LU}]$ into 20 subintervals. The numbers of monomers were counted in every cell of this grid and normalized by $r^2 \sin \theta$. The spatial distribution of the concentration around the bead in the initial state is shown in Fig. 5 (b), while a typical distribution of the concentration during bead motion is displayed in Fig. 5 (c). These results show that before the force was applied to the bead the numbers of monomers in front of the bead and behind it were approximately equal (Fig. 5 (a, b)). This is a consequence of the initial homogeneous distribution of polymers. As soon as the force was applied however, the number of monomer neighbors in front of the bead increased with time while those behind it decreased. For $t\tilde{>}2$ (i.e. after more than $4\times10^4$ steps) both numbers reached steady state values and fluctuated around them during the rest of the simulation. The distribution is characterized by a higher polymer concentration in front of the bead compared to that far from the bead (Fig. 5 a-c)

This inhomogeneous distribution results in an osmotic force resisting the motion of the bead, which is discussed in the following Subsection.

### 4.3. Forces resisting the moving bead

### 4.3.1. Resistive force exerted on the bead by water and by polymers

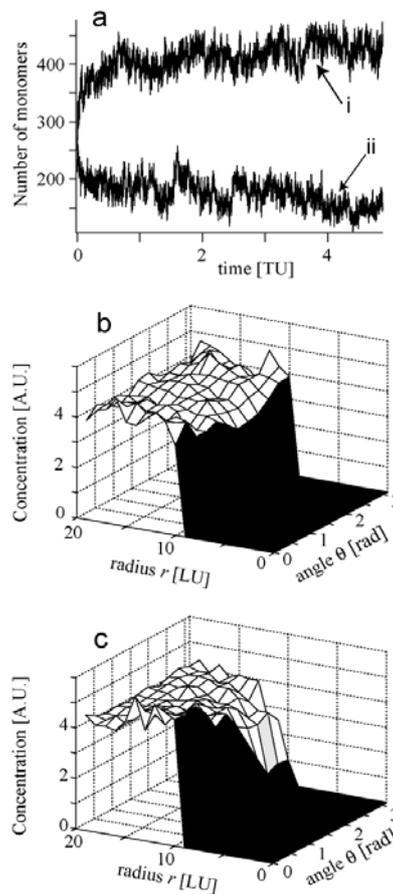

Fig. 5. (a) Number of monomers neighboring the bead at the front (i) and at the rear (ii) hemispheres *versus* time during the bead motion. The external force is applied at $t = 0$. (b, c) The distribution of the concentration of monomers (arbitrary units) in the vicinity of the bead as a function of distance $r$ from the bead center and the azimuthal angle $\theta$. The concentration is shown in the rest state (b) and during the bead motion (c). The moment of time for which the image (c) has been constructed corresponds to the square-root regime of the bead motion.

Before introducing polymers into the system, we checked that the motion of the bead moving through pure water was characteristic of motion in a viscous medium. Details are given in Appendix D.

Computer simulation permitted the direct monitoring of the forces exerted on the bead by the water and the polymers. Denote by $F_w$ the projection of the total force exerted on the bead by the neighboring water spheres in the direction $Ox$ while $F_{pol}$ is the $Ox$ projection of the total force exerted by the neighboring polymers on the bead. These forces are shown in Fig. 6. The external force applied to the bead in this simulation was $f = 1000\text{FU}$ and the mesh size was $\xi \approx 1.57\text{LU}$.

Initially, the absolute value of $F_w$ (Fig. 6 (a)) is close to the external force ($-1000\text{FU}$) but its absolute value decreases with time (Fig. 6 b). By contrast, the force exerted by the polymers on the bead, $F_{pol}$, is approximately zero at the initial moment of motion (Fig. 6 c), a consequence of the initial uniform distribution of the polymers. This force however, gradually grows with time until on average it becomes equal to the external force in its absolute value.

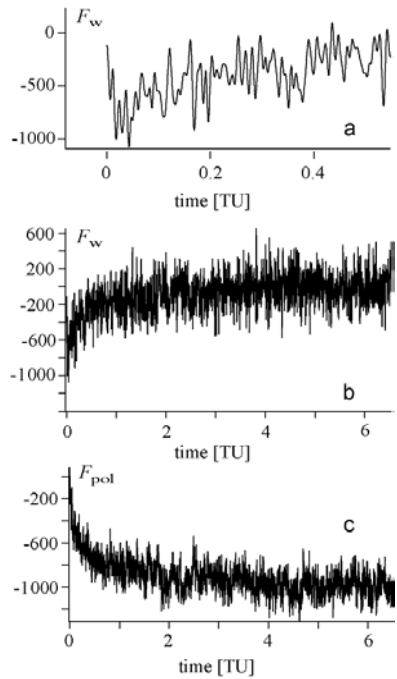

Fig. 6. The forces exerted on the moving bead by water $F_w$ (a) and (b), and by polymers $F_{pol}$ (c). (a) The force due to water, acting on the bead at the beginning of bead motion gives the main contribution to the resistance force. (b) At later times the water contribution to the resistance gradually decreases. (c) The force exerted by polymers on the bead gradually increases with time until its absolute value reaches that of the external force (which is in the shown case $f = 1000\text{FU}$).

### 4.3.2. Comparison of forces exerted by polymers on the bead

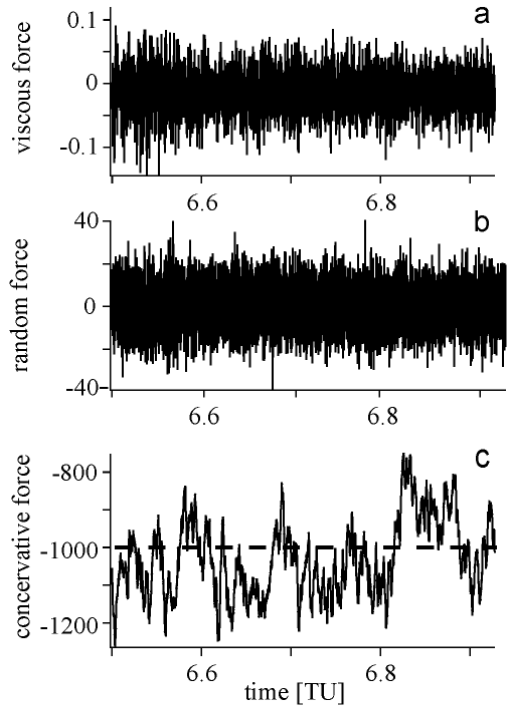

Fig. 7. Three contributions to the force exerted on the bead by the polymers: (a) the viscous, (b) the random and (c) the conservative force (in force units used here). The time interval $6.5\text{TU} \leq t \leq 6.9\text{TU}$ corresponds to the square-root regime. The dashed line in (c) indicates the force equal to $-1000\text{FU}$.

The force exerted by the polymers on the bead consists of three contributions: the viscous, the random and the conservative forces as described by (5)-(7). These forces are shown in Fig. 7 for the time interval $6.5 \leq t \leq 6.9$ which corresponds to the square-root regime. One can see that the conservative force $F_{pol}^{(C)}$ (Fig. 7 (c) is about 50 times larger in its absolute value than the random force (Fig. 7 b) and about four orders of magnitude larger than the viscous friction force Fig. 7 (a).

Note that the absolute value of the viscous force acting on the bead depends on the value of the constant $\gamma_{bm}$ of the viscous interaction between the bead and the monomers which in our work was assigned arbitrarily to be $\gamma_{bm} = 0.1$. Increasing of this constant will increase the contribution of the viscous friction of the polymers to the total resistance force. However, one should expect a linear dependence of the viscous friction force on the constant $\gamma_{bm}$ (compare with Eq. (14) and (16) of the Appendix C). To make the contribution of the dissipative force comparable to that of the conservative force one needs to increase $\gamma_{bm} \sim 10^3$, which is three orders of magnitude larger than $\gamma_{mm}$ and $\gamma_{ww}$. Such a choice seems to be unrealistic. Hence, the conservative force makes the main contribution to the force resisting the bead motion.

The conservative force defined according to (5) represents the steric repulsion between the monomers and the bead taking place during their collisions. The force $F_{pol}^{(C)}$ thus, originates from a pressure difference between the front and rear hemispheres of the bead. The numbers of monomers which are neighbors to the bead at any given time (Fig. 5 a) correlate to $F_{pol}(t)$ (Fig. 6 c). This enables us to deduce that the pressure giving rise to $F_{pol}^{(C)}$ is related to the local entropy decrease due to the piling up of the polymers in front of the moving bead. In this case the pressure can be viewed as the transient osmotic pressure of the polymers and we will refer the force $F_{pol}$ to as the "osmotic force".

Thus, at the initial stage, the bead motion is dominated by the viscosity of the water. This takes place at those times when the distribution of polymers around the bead is relatively uniform (Fig. 5). It is followed by the square-root regime in which the resistance is dominated by the osmotic force. In this regime, the magnitude of the latter is, on average, close to the externally applied force.

**4.4. Diffusion of polymers**

**4.4.1. Free diffusion of polymers**

We have seen that polymers become piled up in a clump in front of the bead during its motion and the resistance must depend on the type of motion of those polymers belonging to the clump. In order to characterize this motion additional simulations were performed.

First we simulated the free diffusion of polymers in the absence of the bead. This enabled us to obtain the transverse, $< R_\perp^2(t) >$, and longitudinal, $< R_\parallel^2(t) >$, mean square displacements (MSD) of the polymers (Appendix D). The former enables one to define the tube diameter. Making use of the relation $\left\langle R_\parallel^2(t) \right\rangle = 2 D_\parallel t$ one finds the longitudinal diffusion coefficient $D_\parallel$.

Figure 8 (a) shows the longitudinal and transverse MSDs of polymers with the same contour length ($L = 34.5\,\text{LU}$) and at various mesh sizes of the network One can see that the longitudinal MSD is almost insensitive to the mesh size while the value of $< R_\perp^2(t) >$ increases with $\xi$. Out of these data the dependence of the average radius of the single-polymer fluctuation tube on the mesh size, $\xi$, was extracted (Fig. 8 b). It exhibits the linear dependence on the mesh size. Fig. 8 (c) shows the dependence of the MSD of polymers on time for polymers of different lengths keeping the mesh size constant. One can see that the transverse MSD is insensitive to polymer length, while the longitudinal MSD decreases with increasing $L$ (shown by the dashed arrow in Fig. 8 c). Our simulation yields the relation $D_\parallel \sim L^{-1}$ (not shown) in accord with the expected dependence of the diffusion coefficient of semiflexible polymers[43].

**4.4.2 Diffusion of polymers in front of the bead**

We analyzed the motion of polymers along the *x*-axis (the direction of bead movement) in front of the moving bead at the time $t = 6.5\,\text{TU}$, which corresponds to the square-root regime. We tracked and analyzed the motion along the x-axis of all monomers which, at $t = 6.5\,\text{TU}$, were situated within a cylindrical domain, coaxial with the bead and in front of it. The radius of the cylinder was equal to that of the bead, while its length was 4LU which is comparable to the distance traveled by the bead from the beginning of the simulation (Fig. 9 (a)). The average displacement of the tracked monomers, which took the form $<\delta x> = \langle x(t-6.5) - x(t=6.5) \rangle$, could be fitted by the expression $<\delta x> \approx Ct^{1/2}$ (Fig. 9 (b)) and yielded $C \approx 0.167$. This suggested that the motion of the monomers in front of the bead and in the direction of bead motion is diffusive with the diffusion coefficient being $D = C^2 / 2$ so that $D \approx 0.014$. This value is close to the longitudinal diffusion coefficient $D_\parallel \approx 0.032$ of polymers of the same length in a solution with the same concentration obtained as described in the previous Section. The same type of analysis as performed in the previous section for the study of longitudinal and transversal diffusion of polymers in solutions was used to study the motion of the segments of polymers formed by the tracked monomers in front of the bead. The results for the longitudinal and transverse MSDs of segments of polymers in front of the bead are shown in Fig. 9 (c). For comparison, the data for the bulk diffusion of polymers are shown on the same graph. One can see that the transverse MSD saturates at distances corresponding to the mesh size, while the longitudinal MSD at $t > 0.2$ exhibits the linear regime $\left\langle R_{\parallel}^2 \right\rangle \sim t$ in both the case of the bulk polymers and of the polymers in front of the bead (Fig. 9 (c)). The good agreement between the two data sets allows one to infer that the motion of the polymers in front of the bead is similar to longitudinal diffusion in the bulk.

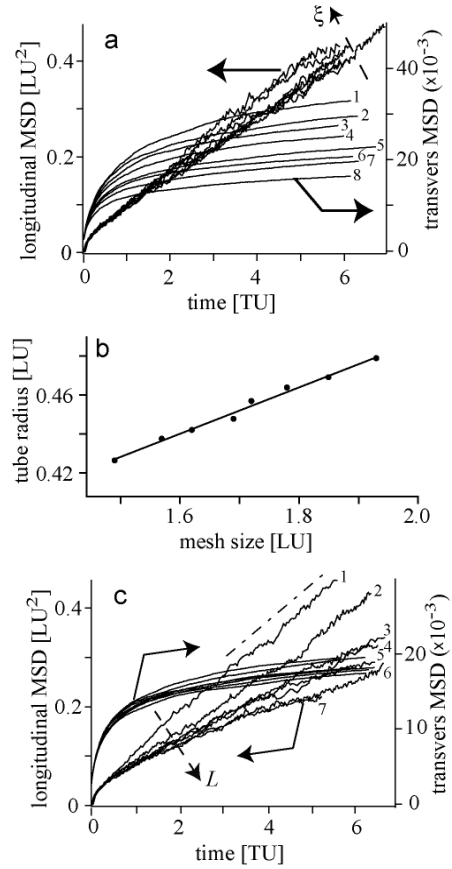

Fig. 8. (a) The longitudinal $<R_\parallel^2>$ (left axis) and transverse $<R_\perp^2>$ (right axis) MSDs of polymers with constant length ($L$=34.5LU) in networks with various mesh sizes. The solid arrows indicate the axis corresponding to the data. The dashed arrow shows the direction of increasing mesh size, $\xi$. (b) The radius of the fluctuation tube as a function of mesh size. Dots show the simulation data, while the solid line is a linear fit. (c) The longitudinal (left axis) and transverse (right axis) MSD of polymers with various contour lengths in a solution with a constant mesh size $\xi \approx 1.57$ LU. The solid arrows indicate the correspondence between the data and the axes. The dashed arrow shows the direction of increasing polymer contour length $L$. The dashed-pointed line shows the slope with $2D_\parallel \approx 0.077$.

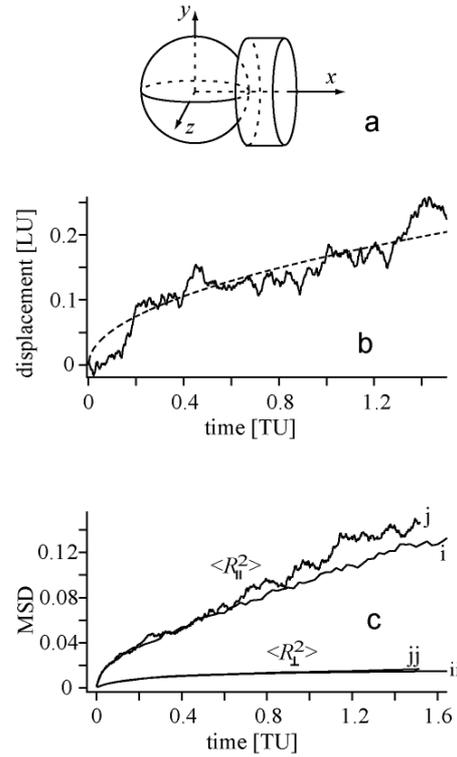

Fig. 9. Motion of polymers located at time, $t = 6.5$ in a cylindrical region, coaxial with the direction of bead motion, the $Ox$ axis, in front of the moving bead (a). (b) Displacement of the monomers. The solid line shows the average displacement $\delta x = \delta x(t)$ of the monomers while the dashed line shows the fit to the data with $\delta x \approx C t^{1/2}$ with $C = 0.167$. (c) The longitudinal (j) and transverse (jj) MSDs of segments of polymers formed by the monomers in front of the bead. For comparison, the longitudinal (i) and transverse (ii) MSD of polymers in the bulk is shown.

Our results show that the polymers pile up in front of the bead as the bead moves. In other words, as the bead moves, a local compression takes place in front of the bead which results in a decrease of the local mesh size (or the radius of the reptation tube) compared to the mesh size far from the bead. The non-zero mean value of the displacement of the monomers $<\delta x>$ describes this compression. The dynamics of the compression is mediated by diffusive motion. *A priori* one might have expected that it takes place both by transverse and by longitudinal diffusion. Our simulations show, however, that the average diffusion coefficient of the monomers in front of the bead is close to the *longitudinal* diffusion coefficient of the free polymers. For this reason it is plausible that the compression process is mainly controlled by longitudinal diffusion.

## 5. Conclusions

We used dissipative particle dynamics to model a bead moving through a solution of entangled semiflexible actin filaments under an applied constant force. In our simulations we accounted for the viscous and random interactions between the three constituents, bead, polymers and water. In addition, the polymer-polymer as well as the bead-polymer interaction included a steric repulsion. In our systems there was no attraction either between the polymers, or between the bead and the polymers. We varied (a) the force applied to the bead, (b) the concentration of the polymers in the solution and (c) the diffusion coefficient of the polymers.

The responses of the bead clearly show two different regimes of bead motion. During the initial regime the response of the bead exhibits a power law $x \sim t^{0.75}$, while in the subsequent regime it obeys the dependence $x \sim t^{\alpha}$ with $\alpha \approx 0.5$.

We found a linear dependence of the response on the applied force, $x \sim f$, a square-root dependence on the longitudinal diffusion coefficient $K \sim D_{\parallel}^{1/2}$ and a power law dependence on the polymer concentration, $K \sim c^{-\gamma_2}$, with $\gamma_2 \approx 1.4$.

We established that the polymers are piled up in front of the bead, while behind the bead the fluid is almost free of polymers. We analyzed the force resisting the motion of the bead and established that, in the square-root regime, the resistance is dominated by the steric repulsion of the polymers related to the osmotic pressure caused by the spatial inhomogeneity of the polymers.

We analyzed the diffusive motion of the polymers in the bulk and compared it with the motion of the polymers in front of the bead. We found that the motion of polymers situated in front of the bead (in the direction of bead motion) is characterized by a diffusion coefficient, which is close to longitudinal diffusion coefficient describing the free diffusion of the polymers in the bulk.

### Appendix A. Constitutive equations of DPD

We let Greek indices $\alpha$ and $\beta$ denote particle type (water, monomer or paramagnetic bead) while indices $i$ and $j$ denote the particular of a given type. In DPD, for $r_{\alpha\beta ij} < R_\alpha + R_\beta$, the conservative repulsion, $\mathbf{F}_{\alpha\beta ij}^{(C)}$, dissipative, $\mathbf{F}_{\alpha\beta ij}^{(D)}$, and random, $\mathbf{F}_{\alpha\beta ij}^{(R)}$, forces acting on particle $\alpha i$ by particle $\beta j$ are defined by

$$\mathbf{F}_{\alpha\beta ij}^{(C)} = a_{\alpha\beta}\left(1 - \frac{r_{\alpha\beta ij}}{R_\alpha + R_\beta}\right)\hat{\mathbf{r}}_{\alpha\beta ij} \tag{5}$$

$$\mathbf{F}_{\alpha\beta ij}^{(D)} = -\gamma_{\alpha\beta} w^{(D)}\left(\frac{r_{\alpha\beta ij}}{R_\alpha + R_\beta}\right)(\hat{\mathbf{r}}_{\alpha\beta ij} \cdot \mathbf{v}_{\alpha\beta ij})\hat{\mathbf{r}}_{\alpha\beta ij} \tag{6}$$

$$\mathbf{F}_{\alpha\beta ij}^{(R)} = \sigma_{\alpha\beta} w^{(R)}\left(\frac{r_{\alpha\beta ij}}{R_\alpha + R_\beta}\right)\theta_{\alpha\beta ij}\hat{\mathbf{r}}_{\alpha\beta ij} \tag{7}$$

If $r_{\alpha\beta ij} \geq R_\alpha + R_\beta$ all the forces $\mathbf{F}_{\alpha\beta ij}^{(C)}$, $\mathbf{F}_{\alpha\beta ij}^{(D)}$ and $\mathbf{F}_{\alpha\beta ij}^{(R)}$ are defined to be zero[50]. Here, $a_{\alpha\beta}$ is the maximum repulsion between the particles of types $\alpha$ and $\beta$, $\mathbf{r}_{\alpha\beta ij} = \mathbf{r}_{\alpha i} - \mathbf{r}_{\beta j}$, $r_{\alpha\beta ij} = |\mathbf{r}_{\alpha\beta ij}|$ and

$\hat{\mathbf{r}}_{\alpha\beta ij} = \mathbf{r}_{\alpha\beta ij} / |\mathbf{r}_{\alpha\beta ij}|$.

$$w^{(D)}\left(\frac{r_{\alpha\beta ij}}{R_\alpha + R_\beta}\right) = \left[w^{R}\left(\frac{r_{\alpha\beta ij}}{R_\alpha + R_\beta}\right)\right]^2 = \left(1 - \frac{r_{\alpha\beta ij}}{R_\alpha + R_\beta}\right)^2 \tag{8}$$

at $r_{\alpha\beta ij} / \left(R_\alpha + R_\beta\right) < 1$ and $w^{(D)} = 0$ otherwise.

$$\sigma_{\alpha\beta}^2 = 2\gamma_{\alpha\beta}k_{\mathrm{B}}T \tag{9}$$

Here $\mathbf{v}_{\alpha\beta ij} = \mathbf{v}_{\alpha i} - \mathbf{v}_{\beta j}$ and $\theta_{\alpha\beta ij}(t)$ is a randomly fluctuating variable with Gaussian statistics:

$$\begin{cases} \left\langle \theta_{\alpha\beta ij}(t) \right\rangle = 0 \\ \left\langle \theta_{\alpha\beta ij}(t)\theta_{\alpha'\beta'i'j'}(t') \right\rangle = (\delta_{\alpha\alpha'}\delta_{ii'}\delta_{\beta\beta'}\delta_{jj'} + \delta_{\alpha\beta'}\delta_{ij'}\delta_{\beta\alpha'}\delta_{ji'})\delta(t-t'). \end{cases} \tag{10}$$

The conservative, dissipative and random forces act along the line connecting the centers of the particles and conserve linear and angular momentum. Using the modified velocity-Verlet algorithm to advance the set of positions and velocities of the particles[55, 62], the random parameter $\theta_{\alpha\beta ij}$ takes the form[55]:

$$\theta_{\alpha\beta ij} = \zeta_{\alpha\beta ij}\Delta t^{-1/2} \tag{11}$$

where $\Delta t$ is the time step of the iteration and $\zeta_{\alpha\beta ij}$ is a dimensionless random number with zero mean and unit variance chosen independently for each pair of interacting particles at each time step.

The time evolution of each particle position obeys Newton's equations of motion:

$$d\mathbf{r}_{\alpha i} / dt = \mathbf{v}_{\alpha i}; \quad m_\alpha d\mathbf{v}_{\alpha i} / dt = \mathbf{f}_{\alpha i} \tag{12}$$

where the subscripts label the $i$-th particle of the type $\alpha$.

$$\mathbf{f}_{\alpha i} = \sum_{\beta, \, j}(\mathbf{F}_{\alpha\beta ij}^{(C)} + \mathbf{F}_{\alpha\beta ij}^{(D)} + \mathbf{F}_{\alpha\beta ij}^{(R)}) \tag{13}$$

where $(\beta, \, j) \neq (\alpha, \, i)$.

## Appendix B. Parameters of the System

The simulations were performed in a box with dimensions $L_y = L_z = 40\,\mathrm{LU}$ and $L_x = 80 \, \text{LU}$. Water spheres possessed radius $R_w = 0.8 \, \text{LU}$, while monomer spheres possessed radius $R_m = 0.3 \, \text{LU}$. The radius of the bead was $R_b = 10 \, \text{LU}$. With this choice of $R_b$ the bead is several times larger than the polymer mesh size and a few times smaller than the length of the polymers corresponding to the experimental conditions[6, 7, 11, 33, 38, 39].

Each polymer was composed of $N_{m/p}$ monomers connected by massless harmonic springs acting between adjacent monomers according to (2). The number of monomers per polymer was chosen $N_{m/p} = 70$. The equilibrium bond length was chosen to be $d = 0.5 \, \text{LU}$, so that, in equilibrium, the polymer contour length was equal to $L = 34.5 \, \text{LU}$. The bond length, $d$, was shorter than the diameter of the monomers, which ensured that two polymer strands did not cross each other, thus providing the topological constraints required by a model of a polymer.

The spring constant was chosen to be $k = 400 \, \text{FU/LU}$. This choice fulfills the requirements that the spring constant value is sufficiently high not to allow the bonds to stretch so much that two polymers would be able to cross one another. Our simulations showed that a crossing occurred of approximately one bond per time unit.

The magnitude of the persistence forces, equation (3), were defined by a choice of the persistent parameter $\mu = 385 \, \text{FU} \times \text{LU}$. We carried out independent simulations to confirm that the polymer bond-bond correlation functions indeed decayed according to $\langle \mathbf{t}_i \cdot \mathbf{t}_j \rangle = \exp\left(-|i-j| \cdot d / L_p\right)$, where $\mathbf{t}_i$ is a unit vector in the direction of the bond linking the monomers labeled $i$ and $i+1$, and that the value of $\mu$ yielded a persistence length of $L_p \approx 150 \, \text{LU}$. The ratio of the persistence length to the bead radius is thus $L_p / R_b = 15$. This is comparable with the value $L_p / R_b \approx 7.6$ appropriate to the experiments[6, 7, 11, 33, 38, 39].

To represent water, we used parameters established by others. The ratio, $\rho$, of the number of the water spheres inside the simulation box to its volume was chosen to be $\rho = 3 \, \text{LU}^{-3}$ (i.e. 3 water spheres per cube of a unit volume[55]).

The parameters $a_{\alpha\beta}$ of the conservative forces and $\gamma_{\alpha\beta}$ of the dissipative forces have the dimensions of FU and FU×TU/LU respectively. For the sake of brevity in the following we often omit these dimensions. The parameter of the water-water conservative force was chosen to be $a_{ww} = 45$. This choice of $\rho$ and $a_{ww}$ ensures that the compressibility of the DPD fluid is close to that of water[55]. We chose the dissipative force constant of the water-water interaction $\gamma_{ww} = 1$. The force constant describing the conservative interaction between the monomers was taken to be of the order of that between waters $a_{mm} = 35$, and we used the constant of the dissipative interaction between the monomers to be $\gamma_{mm} = 5$. For the constant of the monomer-water conservative and dissipative forces we used $a_{mw} = 0$ and $\gamma_{mw} = 1$ respectively. To justify choosing $a_{mw} = 0$ we note that in the framework of the DPD model the water spheres should be viewed as clusters of many individual water molecules. Water can, however, flow past monomers even in regions of high monomer density. This is accounted for by permitting the water clusters to flow through the monomer spheres. The force constants describing the interaction between the bead and the water spheres, as well as between the bead and the monomers were chosen to be $a_{bw} = a_{bm} = 550$ and $\gamma_{bw} = \gamma_{bm} = 0.1$. This choice ensured that with the required water density and water-water repulsion parameter, water penetrated only slightly into the bead. The bead-monomer interaction parameters were taken to be the same as the bead-water ones.

Since inertial effects are irrelevant during the microrheology experiments[6, 7, 11, 33, 38, 39], the results of the simulation must be independent of the masses. For this reason, we have chosen the values of the masses $m_b = 1$ and $m_w = m_m = 10^{-4}$ ensuring that the inertial effects manifest themselves only during the time below 0.01 TU and then vanish. The mass dimension, MU, is obviously related to the other units MU=FU×TU$^2$/LU. The values of the parameters are summarized in Table 1.

The external force acting on the bead was applied along the $Ox$ direction. We found that a suitable simulation time step, $\Delta t$, which provided good temperature control was $5 \times 10^{-5}$ TU. The temperature was controlled by monitoring the mean kinetic energy of the monomer and water spheres, which was, within an accuracy of 2%, equal to $3k_B T / 2$. In all simulations we choose $k_B T = 1$ FU×LU. The thickness of the non-slip layer was chosen to be $\delta_{NS} = 1$ LU.

With the no-slip boundary condition, in which the velocities parallel to those boundaries lying parallel to the direction of bead motion was set equal to zero, it is clear that water particles lose kinetic energy at the boundaries. The average kinetic energy of molecules, was monitored and found to fluctuate around a constant average value $E_{kin} \approx 1.45 k_B T$ which is close to the value $3k_B T / 2$ showing that there was no cooling of the system as a whole despite the loss of kinetic energy at the boundary.

In all simulations the total linear momentum fluctuates in the vicinity of zero in the absence of an external force.

We varied some parameters (e.g. water and monomer radii and interaction constants) in order to check that our principal results were not due to a specific choice. For all parameters used we found the power law $x(t) \sim t^{1/2}$.

### Appendix C. Motion of a Bead in Pure Water

The solution of the Stokes equation[60] yields the viscous friction force

$$f_v = 6\pi R_b \eta v \qquad (14)$$

acting on a bead moving with a constant velocity $v$ through the fluid with the viscosity $\eta$. Within reasonable accuracy the drag force during non-steady motion of the bead could be described by (14) yielding the equation of motion, $m_b dv(t)/dt + 6\pi R_b \eta v(t) = f$. Using the initial condition $v(0) = 0$ one finds the bead velocity:

$$v(t) = f \left[ 1 - \exp\left( -t / \tau \right) \right] / 6\pi R_b \eta \qquad (15)$$

where $\tau = m_\mathrm{b} / 6\pi R_\mathrm{b} \eta$.

Here we compare (15), as well as the velocity field of water calculated analytically[60], with the results of our simulation of the motion of a probe bead in pure water (i.e. in water without polymers).

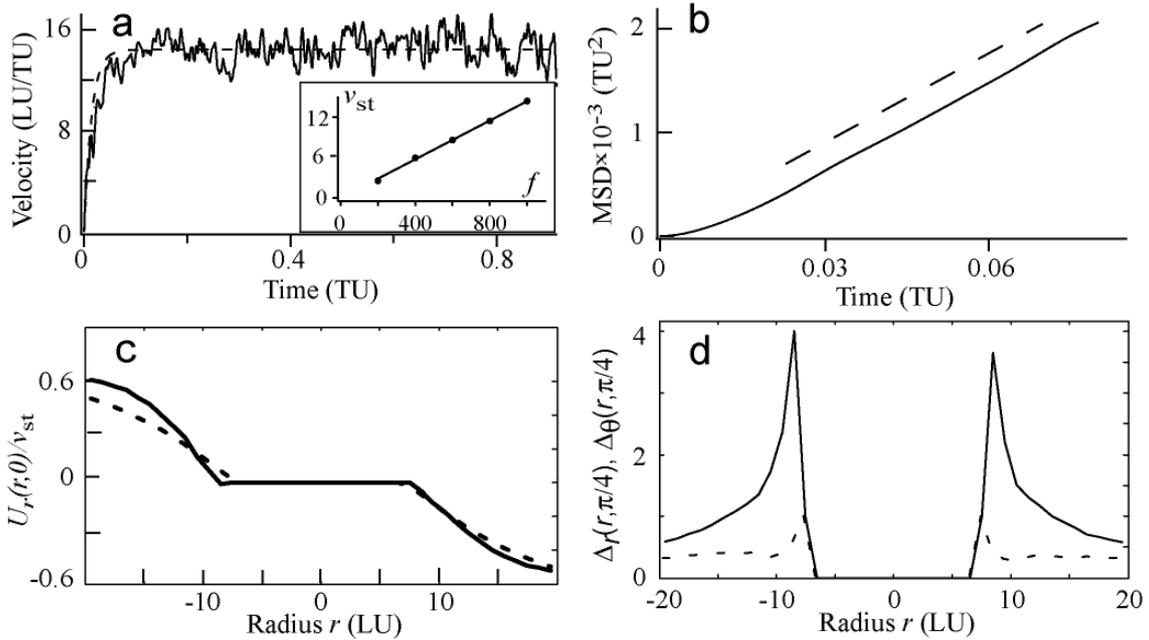

Fig. 10. Bead motion in pure water. (a) Velocity of the bead moving under a constant applied external force *versus* time. The inset shows the dependence of the velocity on the force in the regime of steady motion. (b) *Ox* projection of the mean square displacement of the bead during applied-force-free Brownian motion. The linear relation $< x^2(t) > = 0.0289t$ is shown by the dashed line. (c) The radial component of the water velocity, $\mathbf{U}$, in front and behind the bead ($\theta = 0$, $\theta = \pi$) normalized by the bead velocity $v_\mathrm{st}$ in the steady regime. The results of our simulation are shown by the solid line and the analytical results[60] by the dashed line. (d) The radial and angular functions $\Delta_r (r, \theta)$ (solid line) and $\Delta_\theta (r, \theta)$ (dashed line ) represent measures of deviations of the simulated velocities from the analytical solution of the Stokes equation[60]. They are shown at the section $\theta = \pi / 4$ where they are maximum.

In this simulation the force $f = 1000$ FU was applied to the bead at $t = 0$. The velocity of the bead in pure water is shown in Fig. 10 (a). Fitting (15) to it (the dashed line in Fig. 10 (a)) yields the characteristic time $\tau \approx 0.0143$ TU. At times $t > \tau$ the bead reaches, on average, the regime of steady motion. Since $\tau$ is two orders of magnitude shorter than the typical simulation time, we will not consider inertial effects further. The inset in Fig. 10 (a) shows the dependence of the bead velocity in the steady regime, $v_{st}$, on the applied force, $f$, as well as the linear fit of the data by Eq. (14) which yields the viscosity $\eta \approx 0.37$ FU×TU/LU$^2$.

Alternatively one can extract the viscosity from the analysis of the mean square displacement (MSD) of the bead during its free Brownian motion according to the relation $<x^2(t)> = k_B T t / 3 \pi R_b \eta$. The bead MSD in the $Ox$ direction is shown in the Fig. 10 (b). Fitting the slope (the dashed line in Fig. 10 (b)) one finds $\eta \approx 0.366$ FU×TU/LU$^2$ in a good agreement with our previous result. This agreement follows from satisfying the fluctuation-dissipation theorem.

The value of the viscosity coefficient in DPD was derived in several works[53, 55]. Making use of Eq. (A4) and (A8) of the latter paper[55] one finds $\eta = \eta^{(D)} + \eta^{(K)}$, where

$$\eta^{(D)} = \frac{64 \pi \gamma_{ww} \, \rho^2 R_w^5}{1575}; \quad \eta^{(K)} \approx \frac{45 m_w k_B T}{16 \pi \gamma_{ww} R_w^3} \qquad (16)$$

Substituting $\gamma_{ww} = 1$, $\rho = 3$, $R_w = 0.8$ and $k_B T = 1$ corresponding to the choice of this work one finds $\eta^{(K)} \approx 1.7 \times 10^{-4} << \eta^{(D)}$ and $\eta \approx \eta^{(D)} \approx 0.376$ close to the water viscosity value obtained in our simulation of the bead motion in pure water.

To characterize the water flow around the bead we calculated the velocities, $\mathbf{U}$, of water spheres in the simulation box. In order to compare the velocity field with the known solution of Stokes equation[60] it is convenient to analyze the spherical projections, $U_r(r, \theta)$ and $U_\theta(r, \theta)$, of the velocity field, where the coordinate frame is bound to the bead and the angle $\theta$ is measured from the $Ox$ direction, the direction in which the external force is applied (Fig. 9 a). We took the data when the bead was in the steady regime of motion, and the instant velocity field was smoothed by averaging over 0.5 TU.

Figure 10 (c) shows the radial projection of the water velocity at the line $\theta = 0$, $U_r(r,0)$, normalized by the velocity $v_{st}$ of the steady motion of the bead (solid line in Fig. 10 c). To compare this result with the analytic results[60] we used $R_b = 7$ LU, in order to account for penetration of water into the bead. The ratio $U_r^{(an)}(r,0)/v_{st}$ of the analytic solution of Stokes equation at $\theta = 0$ is shown in Fig. 10 (c) by the dashed line. This comparison shows that along the $Ox$ axis the difference between the simulated and analytical solutions is much smaller than $v_{st}$.

Functions $\Delta_i(r,\theta) = \left| \left( U_i(r,\theta) - U_i^{(an)}(r,\theta) \right) \middle/ U_i^{(an)}(r,\theta) \right|$ (where $i = r$, $\theta$) can be used as measures of deviations of the radial ($i = r$) and the angular ($i = \theta$) components of the velocity field obtained by the simulation, $\mathbf{U}(r,\theta)$, and from the analytical solution $\mathbf{U}^{(an)}(r,\theta)$ of the Stokes equation[60]. Figure 10 (d) shows sections of the functions $\Delta_r(r,\theta)$ (solid line) and $\Delta_\theta(r,\theta)$ (dashed line) taken at $\theta = \pi/4$ at which the deviations are largest. In general, the results of DPD are expected to be reliable on a scale much larger than the size of a water sphere. Indeed, one finds $\Delta_r(r,\pi/4) \geq 1$ only within a distance of a few $R_w$ from the bead surface and vanishes rapidly with increasing radius (Fig. 10 d). The angular deviation, $\Delta_\theta(r,\pi/4)$, is everywhere smaller than unity (Fig. 10 d). Thus the water flow around the bead obtained by our simulations is close to that predicted by the analytical expressions[60].

These results support the simplified approach used in our simulations in which we model the bead as a DPD particle of a large radius, rather than a sphere with non-slip boundary conditions at its surface.

### Appendix D: Simulation of a free diffusion of polymers

In this case, we simulated polymer solutions without a bead. The size of the simulation box was 40LU in all directions. We used periodic boundary conditions and no-slip boundary layers were not applied.

The following procedure was used to study the transverse and longitudinal diffusion coefficients of semiflexible polymers. In the beginning of each simulation, a certain number of cross-sectioning planes were introduced for every polymer. This number varied from 4 to 15 depending on the length of the polymer. The points of cross-sections were chosen to be every $10^{th}$ monomer of a polymer. The cross-section planes were initially defined to be normal to the lines connecting two neighboring monomers. The total number of cross-sections for each simulation was of the order of $10^4$ thus providing ample statistics. During the simulation the coordinates of the points of intersections of polymers with the cross-section planes (the transverse displacements) were stored as well as the out of plane displacement (the longitudinal displacements) of the monomers.

The transverse displacements of the polymers formed "clouds of points" for every cross-section, which were not necessarily centered around zero, since in the initial moment the polymers were not necessarily situated along the centerlines of their tubes. We first calculated the centers of the clouds for all cross-sections formed by the transverse displacements of polymers within a certain time interval $(0,t)$. All clouds were then combined into one and a histogram of displacements with respect to the distance to the origin was produced. The histogram was fitted by a Gaussian distribution $\sim \exp\left(-r^2 / 2R_\perp^2(t)\right)$, where $r$ is the distance from the center of the cloud and $R_\perp(t)$ is the fitting parameter depending on the duration $t$ of the simulation. A similar procedure was used for the study of longitudinal diffusion - except that no center of cloud determination was needed - and a value for $R_\parallel(t)$ was obtained.

**Acknowledgments**: AB was supported by the grant DFG SA264/28-4. This work was supported by NSERC of Canada under a Discovery Grant to DAP. DAP thanks Julian Shillcock and Reinhard Lipowsky for their hospitality at the MPI-KG where he learned about DPD.